\documentclass[prd,twocolumn,amsmath,amssymb,nofootinbib,preprintnumbers,balancelastpage]{revtex4}
\usepackage{graphicx}
\usepackage{hyperref}
\newcommand{\be}{\begin{equation}}
\newcommand{\ee}{\end{equation}}
\newcommand{\bea}{\begin{eqnarray}}
\newcommand{\eea}{\end{eqnarray}}
\hypersetup{
    pdfnewwindow=true,      
    colorlinks=true,       
    linkcolor=black,          
    citecolor=blue,        
    filecolor=blue,      
    urlcolor=blue           
}
\begin{document}
\title{\Large {\bf{{Low Scale Quark-Lepton Unification}}}}
\author{\large{Pavel Fileviez P\'erez$^{1}$, Mark B. Wise$^{2}$} \\}
\affiliation{ \vspace{0.15cm} $^{1}$Particle and Astro-Particle Physics Division,
Max Planck Institute for Nuclear Physics {\rm{(MPIK)}} \\
Saupfercheckweg 1, 69117 Heidelberg, Germany 
\\ 
$^{2}$California Institute of Technology, Pasadena, CA 91125, USA}
\preprint{}
\begin{abstract}
We investigate the possibility that quarks and leptons are unified at a low energy scale 
much smaller than the grand unified scale. A simple theory for quark-lepton unification 
based on the gauge group $SU(4)_C \otimes SU(2)_L \otimes U(1)_R$ is proposed.  
This theory predicts the existence of scalar leptoquarks which could be produced at the 
Large Hadron Collider. In order to have light neutrinos without fine tuning, their masses 
are generated through the inverse see-saw mechanism.
\end{abstract}
\maketitle
\section{Introduction}
In the Standard Model (SM) there are two types of matter fields, the leptons and quarks fields.
The SM with right handed neutrinos describes all the measured properties of quarks and leptons. 
Grand unified theories based on the gauge groups such as $SU(5)$ or $SO(10)$, provide one 
avenue for unifying the properties of quarks and leptons since quarks and leptons 
are part of the same representation of the gauge group. However, in this case the scale of unification is 
very high $M_{GUT} \sim 10^{15-16}~{\rm GeV}$. 

An appealing approach to quark and lepton unification was proposed by J. Pati and A. Salam 
in Ref.~\cite{Pati:1974yy}. They used an $SU(4)_C$ gauge symmetry and the quarks 
and leptons are together in the fundamental representation of the gauge group. 
In this framework the leptons are the fermions with the fourth color. 
This idea also played a major role in grand unification since the Pati-Salam 
gauge group is the maximal subgroup of $SO(10)$. This theory also predicts 
the existence of right-handed neutrinos needed for the seesaw 
mechanism~\cite{TypeI-1,TypeI-2,TypeI-3,TypeI-4,TypeI-5} of neutrino masses.

In this letter we revisit the idea of quark-lepton (QL) unification based on the Pati and Salam paper where the leptons have the fourth color. 
We find a very simple extension of the SM based on the gauge group, $SU(4)_C \otimes SU(2)_L \otimes U(1)_R,$
where  quarks and leptons are unified in the same representation.  For previous studies of these type of models see for example Ref.~\cite{Smirnov95}. 
In addition to the SM fermions the model contains three right handed singlet fermions needed to generate Majorana neutrino masses through 
the inverse seesaw mechanism. 

Assuming near alignment of the quark and lepton generations the experimental limit on the branching ratio for $K_L^0 \rightarrow \mu^{\pm} e^{\mp}$ implies that the scale of $SU(4)_C$ 
breaking must be greater than $1000~{\rm TeV}$. See for example the studies in Refs.~\cite{Valencia, Smirnov} for the constraints coming from meson decays. 
This theory predicts the existence of vector and scalar leptoquarks. While, the vector leptoquarks must be heavy, 
the scalar leptoquarks could be at the TeV scale and give rise to exotic signatures at the Large Hadron Collider.
In this article we discuss the spectrum of our model and outline its main phenomenological consequences.

In section II we  present the simplest model with quark-lepton unification at a low scale that is consistent with the experimental properties of quarks and leptons.
In section III we discuss the properties of the vector and scalar leptoquarks. Finally, we briefly summarize our main results in section IV.

\section{Quark-Lepton Unification}
In models with QL unification based on the idea that leptons have the fourth color~\cite{Pati:1974yy} the SM quarks and leptons can be unified in the same multiplets, 
$( Q_L, \ \ell_L)$, $(u_R, \ \nu_R)$, and $(d_R, \ e_R)$. Therefore,  naively one finds the following relations between 
quark and  lepton masses
\bea
M_u=M_\nu^D,  \  {\rm{and}} \  M_e=M_d,
\eea 
where $M_u$, $M_\nu^D$, $M_e$ and $M_d$ are 
the up quark, Dirac neutrino, charged lepton and down quark masses, respectively. 
As is well-known, these relations are not consistent with experiment.

 We now construct  the simplest model of quark lepton unification based on the gauge group
\begin{equation}
 G_{QL}=SU(4)_C \otimes SU(2)_L \otimes U(1)_R,
 \end{equation}
that  is consistent with experimental results on the properties of quarks and leptons evading the unacceptable mass relations discussed above and that allows the scale for the symmetry breaking $G_{QL} \rightarrow G_{SM}=SU(3)\otimes SU(2)_L\otimes U(1)_Y$ to be much smaller than the grand unification scale.

The fermion matter fields are in the representations 
\begin{eqnarray}
F_{QL}&=&
\left(
\begin{array}{cc}
u 
&
\nu 
\\
d 
&
e
\end{array}
\right) \sim (4,2,0),
\\
F_u&=&
\left(
\begin{array}{cc}
u^c 
&
\nu^c
\end{array}
\right) \sim (\bar{4},1,-1/2), 
\\ 
F_d&=&
\left(
\begin{array}{cc}
d^c 
&
e^c
\end{array}
\right) \sim (\bar{4},1,1/2).
\end{eqnarray}
Here we have chosen to work only with left-handed fermion fields as is common in discussions of grand unified theories.

The gauge group $G_{QL}$ is spontaneously broken to $G_{SM}$ by the vacuum expectation value of the scalar
field
\begin{eqnarray}
\chi = \left(  \chi_u  \  \ \chi_R^0 \right) \sim (4,1,1/2).
\end{eqnarray}
Without loss of generality the vacuum expectation value can be taken to be only in  the fourth component, $\langle \chi_R^0 \rangle =v_{\chi}/\sqrt{2}$. The SM hypercharge $Y$ is given by,
\begin{equation}
Y=R + {\sqrt{6} \over 3 }T_4,
\end{equation}
where $T_4$ is the properly normalized  $SU(4)_C$ generator, that acting on the fundamental $4$ representation
is the diagonal matrix
\begin{eqnarray}
T_4 &=&
\frac{1}{2 \sqrt{6}}
\left(
\begin{array} {cccc}
1 & 0 & 0 & 0 \\
0 & 1 & 0 & 0 \\
0 & 0 & 1 & 0 \\
0 & 0 & 0 & -3 \\
\end{array}
\right).
\end{eqnarray}
To break the gauge group down to the low energy $SU(3)_C \otimes U(1)_Y$ gauge group in a way that can give acceptable fermion masses we add two more scalar representations,
a  Higgs doublet
\begin{eqnarray}
H^T &=& \left( H^+   \  H^0 \right) \sim (1,2,1/2), 
\end{eqnarray}
and the scalar $\Phi \sim (15,2,1/2)$,
\begin{eqnarray}
\Phi &=&
\left(
\begin{array} {cc}
\Phi_8 & \Phi_3  \\
\Phi_4 & 0  \\
\end{array}
\right) + T_4 \ H_2,
\end{eqnarray}
which contains a second Higgs doublet  $H_2$. The new scalars in $\Phi$ are the second Higgs doublet, the color octet  with the same weak quantum numbers as the Higgs doublet, $\Phi_8 \sim (8,2,1/2)_{SM}$,  studied by Manohar and Wise in Ref.\cite{Manohar:2006ga} and the scalar leptoquarks  
$\Phi_3 \sim (\bar{3},2,-1/6)_{SM}$ and  $\Phi_4 \sim (3,2,7/6)_{SM}$.  For the use of the field $\Phi$ in these type of models see Refs.~\cite{15field}.
These scalar leptoquarks do not give rise to proton decay~\cite{Arnold:2013cva} at the renormalizable level since they do not couple to a quark pair. 
Proton decay occurs at the dimension six level.

The Yukawa interactions in this theory are given by
\begin{eqnarray}
{\cal L}_{QL}^{Y} &=&
Y_1  {F}_{QL} F_u H  \ + \ Y_2  {F}_{QL} F_u \Phi   \nonumber \\
& + &  Y_3 \  H^\dagger {F}_{QL} F_d  \ + \  Y_4  \Phi^\dagger {F}_{QL}  F_d   + \mbox{h.c.},
\end{eqnarray}
which after symmetry breaking give rise to the following mass matrices for the SM fermions 
\begin{eqnarray}
M_u &=& Y_1 \frac{v_1}{\sqrt{2}} + \frac{1}{2 \sqrt{6}} Y_2 \frac{v_2}{\sqrt{2}}, \\
M_\nu^D &=& Y_1 \frac{v_1}{\sqrt{2}} - \frac{3}{2 \sqrt{6}} Y_2 \frac{v_2}{\sqrt{2}}, \\
M_d &=& Y_3 \frac{v_1}{\sqrt{2}} + \frac{1}{2 \sqrt{6}} Y_4 \frac{v_2}{\sqrt{2}}, \\
M_e &=& Y_3 \frac{v_1}{\sqrt{2}} - \frac{3}{2 \sqrt{6}} Y_4 \frac{v_2}{\sqrt{2}}.
\end{eqnarray}
Here the vacuum expectation values (VEVs)  that break  $G_{SM}$ are
\begin{eqnarray}
\left< H^0 \right>  &=& \frac{v_1}{\sqrt{2}},~{\rm{and}} \  \ \left< H^0_2 \right> = \frac{v_2}{\sqrt{2}}.
\end{eqnarray}
Since there are four independent Yukawa coupling matrices in the above equations we can generate acceptable masses for all the quarks and leptons. 
However, in order to achieve light Dirac neutrino masses one needs a {\textit{severe fine tuning}} between the two terms contributing to $M_\nu^D$ in Eq.~(13). 
See Ref.~\cite{Foot} for an alternative model using the Pati-Salam symmetry.

It is useful to note that the renormalizable  couplings of the model contain an automatic global $U(1)$ fermion matter symmetry $U(1)_F$ where the matter charges of the fermion fields  $F_{F_{QL}}=1$, $F_{F_u}=F_{F_d}=-1$. The scalar fields do not transform under this symmetry.

Although this model can be consistent with experiment the fine tuning needed to get very light Dirac neutrinos is not attractive. A modest extension of the fermion content of the model allows us to avoid this fine-tuning.

We can generate small Majorana  masses for the light neutrinos if we add three new singlet left handed fermionic fields $N$ and use the following interaction terms
\begin{eqnarray}
{\cal L}_{QL}^\nu &=&
Y_5 F_u \chi N  \ + \  \frac{1}{2} \mu N N   + \mbox{h.c.}.
\end{eqnarray} 
For simplicity, we now discuss the neutrino sector in the one generation case. The discussion generalizes easily to the three generation case. 
Then, there are three left handed neutrino fields (one each of $\nu$, $\nu^c$ and $N$) and the neutrino mass matrix reads as
\bea
\left( \nu \  \nu^c \  N  \right) 
\left(\begin{array}{ccc} 
0 & M_\nu^D & 0  \\ 
(M_\nu^D)^T & 0 & M_\chi^D \\
0 &  (M_\chi^D)^T & \mu
\end{array}\right)  
\left(\begin{array}{c} \nu \\  \nu^c \\ N  \end{array}\right).
\eea
Here $M_\nu^D$ is given by Eq.(13) and 
\begin{equation}
M_\chi^D = Y_5 \frac{v_\chi}{\sqrt{2}}.
\end{equation}
Assigning the $N$ fermion charge $F_N=1$  only the term proportional to $\mu$ breaks the matter symmetry. 
Hence it is natural to take this parameter to be much smaller than the other entries in the mass matrix. 

In the limit when, $\mu << M_{\nu}^D, M_\chi^D$, the model has a heavy Dirac neutrino with mass, $\sqrt{(M_\chi^D)^2+(M_\nu^D)^2}$, 
where $\nu^c$ is paired with the linear combination
\bea
\nu_h= (M_{\chi}^DN + M_\nu^D \nu)/\sqrt{(M_\chi^D)^2+(M_\nu^D)^2}.
\eea 
The orthogonal linear combination is the light Majorana neutrino
\begin{equation}
\nu_{l}=(M_{\chi}^D\nu- M_\nu^D N)/\sqrt{(M_\chi^D)^2+(M_\nu^D)^2}.
\end{equation}
It would be massless in the limit $\mu \rightarrow 0$ since then Majorana masses are forbidden by the fermion matter symmetry. 

So for this neutrino to have the same properties as in the SM we need  $M_\nu^D <<M_\chi^D$ which is reasonable when the scale of SM symmetry breaking is much 
smaller than the scale of $SU(4)_C\otimes U(1)_R$ symmetry breaking. When the parameters in the neutrino mass matrix follow the relation, $\mu << M_\nu^D << M_\chi$,  the Majorana light neutrino masses is given by
\begin{equation}
m_\nu = \mu  (M_\nu^D)^2 / (M_\chi^D)^2, 
\end{equation}
which is the usual relation in the inverse seesaw mechanism~\cite{IS1,IS2}. Therefore, we can have light neutrinos without fine tuning. 
If $M_\nu^D \sim 10^2$ GeV and $M_\chi^D \sim 10^6$ GeV, the neutrino mass satisfies $m_\nu \sim \mu \times 10^{-8}$. Thus, $\mu$ has 
to be very small, smaller than $0.1$ GeV, but it is protected by the matter symmetry. 

If we use the usual seesaw mechanism the scale for QL unification has to be close to the seesaw scale $\sim 10^{14}$ GeV. See for example Ref.~\cite{Goran}. 
However, the inverse seesaw mechanism allows us to have a low scale for QL unification as we have discussed above.

The $SU(4)_C$ gauge boson, $A_\mu \sim (15,1,0)$, can be written as
\begin{eqnarray}
A_\mu &=&
\left(
\begin{array} {cc}
G_\mu & X_\mu / \sqrt{2}  \\
X_\mu^* / \sqrt{2} & 0  \\
\end{array}
\right) + T_4 \ B_\mu^{'}.
\end{eqnarray}
Here $G_\mu \sim (8,1,0)_{SM}$ are the gluons and $X_\mu \sim (3,1,2/3)_{SM}$ are the new massive vector lepto-quarks. 
The different transformation properties under color of $X_{\mu}$ and $\Phi_3$ arise from our conventions for how the two 15's 
$A_{\mu}$ and $\Phi$ transform under $SU(4)_C$ gauge transformations $U$. Neglecting the space-time dependence of the transformations,  
$A_{\mu}  \rightarrow U A_{\mu} U^{\dagger}$ while $\Phi \rightarrow U^* \Phi U^T$.

We have mentioned before that the Higgs sector is composed of three Higgses, $H \sim (1,2,1/2)$, $\chi \sim (4,1,1/2)$ and $\Phi \sim (15,2,1/2)$.
Therefore, the scalar potential can be written as
\bea
V&\supset& m_H^2 H^\dagger H + m_\chi^2 \chi^\dagger \chi + m_\Phi^2 {\rm{Tr}} (\Phi^\dagger \Phi) + \lambda_1 H^\dagger H \chi^\dagger \chi \nonumber \\
&+& \lambda_2 H^\dagger H  {\rm{Tr}} (\Phi^\dagger \Phi) + \lambda_3 \chi^\dagger \chi  {\rm{Tr}} (\Phi^\dagger \Phi) \nonumber \\
&+& ( \lambda_4 H^\dagger \chi^\dagger \Phi \chi + {\rm{h.c.}} ) + \lambda_5 H^\dagger {\rm{Tr}} (\Phi^\dagger \Phi)  H + \lambda_6 \chi^\dagger \Phi \Phi^\dagger \chi \nonumber \\
&+& \lambda_7 (H^\dagger H)^2 + \lambda_8 (\chi^\dagger \chi)^2 + \lambda_9 {\rm{Tr}} (\Phi^\dagger \Phi)^2 \nonumber \\
&+ & \lambda_{10} ({\rm{Tr}} \Phi^\dagger \Phi)^2.
\eea
Here the trace is only in the $SU(4)_C$ space.
We assume that the parameters of the potential can be chosen so that these additional scalars  get vacuum expectation values that leave color and the electromagnetic charge $Q=T^3_L+Y$ unbroken, {\it i.e.}, only the neutral components of $\chi$, $H$ and $H_2$ get expectation values.
\section{Vector and Scalar Leptoquarks}
This theory predicts the existence of vector and scalar leptoquarks.
The vector leptoquarks, $X_\mu \sim (3,1,2/3)_{SM}$, have the following interactions
\bea
{\cal{L}}& \supset& \frac{g_4}{\sqrt{2}} X_\mu \left(  \bar{Q}_L  \gamma^\mu \ell_L +  \bar{u}_R \gamma^\mu \nu_R +  \bar{d}_R \gamma^\mu e_R  \right) 
 +  \rm{h.c.}. \nonumber \\
\eea 
The gauge coupling $g_{4}$ is equal to the strong coupling constant evaluated at the $SU(4)_C$ scale.
The vector leptoquarks contribute to the rare meson decays, $K_L^0 \to e^\mp \mu^\pm$, which give a lower bound 
on the $SU(4)_C$ scale. This issue has been studied by several groups, see for example Refs.~\cite{Valencia,Smirnov}, 
and the bound is $M_X \geq 10^3$ TeV. Notice that there is freedom in the unknown mixings between quarks and leptons 
and one can have a lower scale. For simplicity we assume that there is no suppression mechanism.

The scalar leptoquarks, $\Phi_3 \sim (\bar{3},2,-1/6)_{SM}$ and $\Phi_4 \sim (3,2,7/6)_{SM}$, present in the field $\Phi \sim (15,2,1/2)$, 
have the following interactions
\bea
{\cal{L}}& \supset& Y_2 Q_L \Phi_3 \nu^c  + Y_2  \ell_L  \Phi_4 u^c \ + \  \nonumber \\ 
&& Y_4 Q_L  \Phi_4^\dagger e^c + Y_4  \ell_L  \Phi_3^\dagger d^c +  {\rm{h.c.}} 
\eea 
It is important to mention that the only coupling constrained by $K_L^0 \to e^\mp \mu^\pm$ is $Y_4$, 
but this coupling is small, below $10^{-2}$, to be in agreement with fermion masses. The coupling 
$Y_2$ is less constrained and can be small as well. Therefore, the scalar leptoquarks 
$\Phi_3$ and $\Phi_4$ can be at the TeV scale and be produced at the Large Hadron Collider.
As is well known, one can have the QCD pair production of leptoquarks and the decays into a jet 
and lepton can give a unique signal. For a recent discussion of the leptoquark signatures at the LHC see Refs.~\cite{Han,Kramer,Davidson,Davidson2}.
A detailed analysis of the constraints coming from meson decays, the lepton number violating decays, and the collider signatures 
is beyond the scope of this letter.
\section{Concluding Remarks}
In this article we have proposed a simple theory where the Standard Model quarks and leptons are unified 
using the gauge symmetry $SU(4)_C \otimes SU(2)_L \otimes U(1)_R$. The neutrinos are 
Majorana fermions and their masses are generated through the inverse seesaw mechanism. The quark and lepton unification scale 
can be as low as $10^3$ TeV. The main constraints on the QL breaking scale are coming from the rare meson decays mediated 
by the vector leptoquark. This theory predicts the existence of scalar 
leptoquarks, $\Phi_3 \sim (\bar{3},2,-1/6)_{SM}$ and $\Phi_4 \sim (3,2,7/6)_{SM}$, which could be at the TeV scale and give rise to exotic signatures at the 
Large Hadron Collider. Subjects for further work include studying the correlation between the collider signals and the different constraints coming 
from flavour violation, and constructing the supersymmetric version of the theory to solve the hierarchy problem. We would like to mention that this 
theory could have an UV completion based on the Pati-Salam gauge group or could arrive from a grand unified theory based on $SU(6)$. 

{\textit{Acknowledgment}}:
P. F. P. thanks the theory group at Caltech for their great hospitality. 
This paper is funded by the Gordon and Betty Moore Foundation through
Grant $\# 776$ to the Caltech Moore Center for Theoretical Cosmology
and Physics. The work of M.B.W. was supported also in part by the
U.S. Department of Energy under contract No. DE-FG02-92ER40701. 


\begin{thebibliography}{000}


\bibitem{Pati:1974yy}
  J.~C.~Pati and A.~Salam,
  ``Lepton Number as the Fourth Color,''
  Phys.\ Rev.\ D {\bf 10} (1974) 275
   [Erratum-ibid.\ D {\bf 11} (1975) 703].
  
\bibitem{TypeI-1}
  P.~Minkowski,
  ``$\mu \to e \gamma$  at a rate of one out of $10^9$ muon decays?,''
  Phys.\ Lett.\ B {\bf 67} (1977) 421.
  
\bibitem{TypeI-2}
  R.~N.~Mohapatra and G.~Senjanovi\'c,
  ``Neutrino Mass and Spontaneous Parity Nonconservation,''
  Phys.\ Rev.\ Lett.\  {\bf 44} (1980) 912.

\bibitem{TypeI-3}
  M.~Gell-Mann, P.~Ramond and R.~Slansky,
   in {\it Supergravity}, eds.\ P.~van Nieuwenhuizen et al.,
   (North-Holland, 1979), p.~315.
   
\bibitem{TypeI-4}   
  S.~L.~Glashow, in {\it Quarks and Leptons}, Carg\`ese, eds.\ M.~L\'evy et al.,
(Plenum, 1980), p.~707.

\bibitem{TypeI-5}
  T.~Yanagida,
{Proceedings of the Workshop on the Unified Theory
  and the Baryon Number in the Universe}, eds.\ O.\ Sawada et al.,
p.~95, KEK Report 79-18, Tsukuba (1979).


\bibitem{Smirnov95}
  A.~D.~Smirnov,
  ``The Minimal quark - lepton symmetry model and the limit on Z-prime mass,''
  Phys.\ Lett.\ B {\bf 346} (1995) 297
  [hep-ph/9503239].

   

\bibitem{Valencia}
  G.~Valencia and S.~Willenbrock,
  ``Quark - lepton unification and rare meson decays,''
  Phys.\ Rev.\ D {\bf 50} (1994) 6843
  [hep-ph/9409201].
  
\bibitem{Smirnov}
  A.~D.~Smirnov,
  ``Mass limits for scalar and gauge leptoquarks from $K_L^0 \to e^{\mp} \mu^\pm$, $B_0 \to e^\pm \tau^\mp$ decays,''
  Mod.\ Phys.\ Lett.\ A {\bf 22} (2007) 2353
  [arXiv:0705.0308 [hep-ph]].
  
  
  \bibitem{Manohar:2006ga} 
  A.~V.~Manohar and M.~B.~Wise,
  ``Flavor changing neutral currents, an extended scalar sector, and the Higgs production rate at the CERN LHC,''
  Phys.\ Rev.\ D {\bf 74}, 035009 (2006)
  [hep-ph/0606172].
  
  
  \bibitem{15field}
  J.~C.~Pati, A.~Salam and U.~Sarkar,
  ``$\Delta B = - \Delta L$, $n \to e^- \pi^+, e^- K^+, \mu^- \pi^+$ and $\mu^- K^+$ decay modes in $SU(2)_L \times SU(2)_R \times SU(4)_C$ or $SO(10)$,''
  Phys.\ Lett.\ B {\bf 133} (1983) 330.

      
  
\bibitem{Arnold:2013cva} 
  J.~M.~Arnold, B.~Fornal and M.~B.~Wise,
  ``Phenomenology of scalar leptoquarks,''
  arXiv:1304.6119 [hep-ph].
  
\bibitem{Foot}
  R.~Foot,
  ``An Alternative SU(4) x SU(2)L x SU(2)R model,''
  Phys.\ Lett.\ B {\bf 420} (1998) 333
  [hep-ph/9708205].
  
  
\bibitem{IS1}
  R.~N.~Mohapatra,
  ``Mechanism For Understanding Small Neutrino Mass In Superstring Theories,''
  Phys.\ Rev.\ Lett.\  {\bf 56} (1986) 561.
  
\bibitem{IS2}
  R.~N.~Mohapatra and J.~W.~F.~Valle,
  ``Neutrino Mass and Baryon Number Nonconservation in Superstring Models,''
  Phys.\ Rev.\ D {\bf 34} (1986) 1642.
  
  
\bibitem{Goran}
  A.~Melfo and G.~Senjanovic,
  ``Minimal supersymmetric Pati-Salam theory: Determination of physical scales,''
  Phys.\ Rev.\ D {\bf 68} (2003) 035013
  [hep-ph/0302216].
  
  
\bibitem{Han}
  P.~Fileviez Perez, T.~Han, T.~Li and M.~J.~Ramsey-Musolf,
  ``Leptoquarks and Neutrino Masses at the LHC,''
  Nucl.\ Phys.\ B {\bf 819} (2009) 139
  [arXiv:0810.4138 [hep-ph]].

\bibitem{Kramer}
  M.~Kramer, T.~Plehn, M.~Spira and P.~M.~Zerwas,
  ``Pair production of scalar leptoquarks at the CERN LHC,''
  Phys.\ Rev.\ D {\bf 71} (2005) 057503
  [hep-ph/0411038].
  

\bibitem{Davidson}
  S.~Davidson and S.~Descotes-Genon,
  ``Minimal Flavour Violation for Leptoquarks,''
  JHEP {\bf 1011} (2010) 073
  [arXiv:1009.1998 [hep-ph]].
  
  
\bibitem{Davidson2}
  S.~Davidson and P.~Verdier,
  ``Leptoquarks decaying to a top quark and a charged lepton at hadron colliders,''
  Phys.\ Rev.\ D {\bf 83} (2011) 115016
  [arXiv:1102.4562 [hep-ph]].
  
  
\end{thebibliography}
\end{document}